# Road to perdition? The effect of illicit drug use on labour market outcomes of prime-age men in Mexico[*]

José-Ignacio Antón[†]

Juan Ponce[‡]

Rafael Muñoz de Bustillo[†]

[†]*University of Salamanca*

[‡]*FLACSO Ecuador*

3rd May 2024

**Abstract**

This study addresses the impact of illicit drug use on the labour market outcomes of men in Mexico. We leverage statistical information from three waves of a comparable national survey and make use of Lewbel's heteroskedasticity-based instrumental variable strategy to deal with the endogeneity of drug consumption. Our results suggests that drug consumption has quite negative effects in the Mexican context: It reduces employment, occupational attainment and formality and raises unemployment of local men. These effects seem larger than those estimated for high-income economies.

**Keywords:** drugs, labour market, employment, unemployment, formality, Mexico.

**JEL classification:** J22, I12.

# 1. Introduction

Over the past two decades, Mexico has unfortunately become infamous as a major international supplier and traffic hub of illicit addictive substances and a battleground between local drug cartels and between those organisations and the state (United Nations Office on Drugs and Crime [UNODC], 2023). This violence has taken a dramatic economic toll, deterring growth, encouraging internal and international migration and harming the functioning of the labour market (Gutiérrez-Romero &

---

[*] Corresponding author: José-Ignacio Antón, Department of Applied Economics, University of Salamanca, Campus Miguel de Unamuno, 37007 Salamanca (Spain), email: janton@usal.es. We thank Patricia Triunfo for helpful comments on an earlier draft.

Oviedo, 2018; Orozco-Alemán & González-Lozano, 2018; Velasquez, 2020). However, despite the dramatic increase in consumption in a highly punitive environment (Rangel Romero, 2023)—the percentage of Mexicans who had taken illicit addictive substances at some point in their lives rose from 5.2% in 2008 to 9.9% in 2016–2017 (Instituto Nacional de Psiquiatría Ramón de la Fuente Muñiz, 2017)—we lack knowledge about the consequences of drug use (other than alcohol or tobacco use) on nationals.

Concerns about the negative effects of consuming drugs stem from the severe impacts on numerous dimensions of users' lives, such as health, education and employment outcomes, and externalities, such as car accidents, criminal activity and an increasing burden on public health care. Despite competing economic theories to explain addictive behaviour (Cawley & Ruhm, 2011) and nontrivial differences between substances, in general, the academic literature predicts that a key channel through which drug use can harm labour market success is productivity losses due to its health consequences (Culyer, 1973), which can result in chronic absenteeism and dismissal, less socially conforming behaviours (Kandel, 1984) and unfavourable signals to employers (Kandel et al., 1995).

This study aims to shed light, for the first time, on the impact of illicit drug use on the labour market performance of prime-age Mexican men.[1] Leveraging three waves of a national survey on drug consumption patterns and making use of the heteroskedasticity-based instrumental variable (IV) approach proposed by Lewbel (2012)—a reasonable alternative when convincing external instruments are unavailable—we explore the effects of drug use on employment, unemployment, occupational attainment and formality.

Our findings suggest that drug consumption exerts a negative effect on employment, raises the likelihood of unemployment and reduces the probability of working in a white-collar job or in the formal sector of the economy. Our main results mostly hold when we consider an alternative (imperfect) IV used in other studies of this type (religious affiliation) and are robust to different classifications of drugs and the exclusion of other potentially endogenous variables (marital status and education) from the control covariates.

The economic literature that aims to disentangle how drug consumption affects labour market outcomes is extensive.[2] Van Ours and Williams (2015) group the prior research on the consequences of drug use on labour market success into two waves. The first one, which dates from the early nineties and focuses on the United States, consists mainly of works that treat drug consumption as exogenous or use instrumental variables whose exogeneity is highly questionable, such as family background characteristics, religiosity, the frequency of going to bars or the experience of a recent divorce. Much of this literature reports rather counterintuitive results associating the use of various drugs with much higher labour force participation and earnings.[3]

---

[1] Until 2021, the only legal drugs in Mexico were alcohol and tobacco. The Mexican Supreme Court of Justice declared the prohibition of marijuana consumption for recreational use unconstitutional and reconvened the national parliament to regulate it, which it did in March 2021. The period analysed in this paper (2008–2016) predates the legalisation of cannabis.
[2] For the sake of brevity, we refrain from including in the discussion any of the numerous works taking a public policy or medicine perspective, which do not address the endogeneity of drug use.

The second wave of studies began in the late nineties, includes other developed countries—the Netherlands, the United Kingdom and Spain—and seeks to improve on the earlier work in one or more of the following ways: determining whether the drug dose is critical to the labour market outcomes, paying attention to timing (e.g., whether the eventual harmful effects of drugs on productivity are immediate or delayed) and developing more convincing identification strategies (more carefully selecting instruments and testing their strength, considering lifetime rather than current use, leveraging the timing of the events or employing structural econometric techniques).[4] The results of this stream of research are far from unanimous, although they tend to suggest that occasional use of drugs or soft addictive substances does little harm, while problematic (intense and continued) consumption and hard drugs could seriously damage labour market prospects.

In addition to the two previous waves mentioned, we can also highlight several recent studies that focus on the opioid crisis in the United States (Aliprantis et al., 2023; Beheshti, 2022; Kaestner & Ziedan, 2023; Harris et al., 2020; Park & Powell, 2021), which leverage quasi-experimental spatial variation due to state-specific public policies or the presence of large prescribers. This research, which benefits from more rigorous identification strategies, finds that opioid use has negative impact of labour market participation and earnings.

Overall, three salient features emerge from the assessment of previous research. First, a significant proportion of earlier works relies on dubious instrumentation strategies, where the exclusion restriction is unlikely to hold. This highlights the difficulty of finding good external instruments. Second, apart from the last stream of studies focused on the American opioid epidemic, the lack of consensus in the results is clear. Last, evidence from countries other than highly developed economies is absent.[5]

---

[3] See, for example, Gill and Michaels (1992), Kaestner (1991, 1994a, 1994b, 1995), Kaestner and Grossman (1995) and Register and Williams (1992).

[4] See, for instance, Buchmueller and Zuvekas (1998), Burgess and Proper (1998), DeSimone (2002), French, Roebuck and Alexander (2001), French, Zarkin and Dunlap (1998), Kandel et al. (1995), MacDonald and Pudney (2000a, 2000b, 2001), van Ours (2004, 2006), Ringel et al. (2006) and Zarkin, Mroz et al. (1998). It is worth mentioning the work of DeSimone (2002), who uses the degree of depenalisation of marijuana by state in the United States and the local price of the same drug as instruments. Willams and Skeels (2006) use a similar instrument in their study on the health effects of cannabis consumption in Australia. Because of their similarities, we can include in this second wave of studies several works not included in van Ours and Williams's (2015) survey, such as Casal et al. (2020), Ringel et al. (2006), Rivera et al. (2013), Casal et al. (2020) and Williams and van Ours (2020). Conti (2010) and Mezza and Buchinsky (2021) are examples of works employing structural econometric methods.

[5] For brevity, we intentionally limit our literature review to works that focus on illicit substances, which are often analysed separately from legal drugs such as alcohol and tobacco. To the best of our knowledge, the only study of this kind for Mexico is that of Suárez Martínez and Caamal-Olvera (2021), who find a positive impact of alcohol consumption on teenage employment.

This work contributes to the literature in two ways. To the best of our knowledge, all previous research has focused on high-income economies, with labour markets that are substantially different from the Mexican one, more generous social protection networks and a wider range of antidrug policies and campaigns. We are therefore the first to provide evidence for a middle-income country. Moreover, we are pioneers in making use of Lewbel's (2012) IV approach based on the heteroskedasticity of residuals, which allows us to deal with simultaneity and measurement error, in the context of addressing the impact of drug consumption.

The rest of the paper unfolds in three sections as follows. Section 2 discusses our research design, providing details on the database and the empirical strategy adopted. Section 4 presents the results of our analyses and discusses their heterogeneity and robustness. Section 5 summarises and discusses the main implications of the research.

## 2. Data and methods

### 2.1. Data

We exploit the statistical information provided by the National Survey of Addictions (NSA) 2008, 2011 and 2016–2017, administered by the Mexican National Institute of Statistics and Geography.[6] The NSA has a four-cluster and stratified sampling design and is representative of the Mexican population aged 12 to 65. It includes a rich set of questions on drug use and initiation patterns and relevant information on both household and individual socioeconomic characteristics. Importantly, the questions on drug use are completely consistent across the three waves.

We make several decisions regarding the sample used in the analyses. First, we focus on men because the prevalence of drug use among women in Mexico is extremely low (less than 0.01% of women aged 22–50 in our sample consumed these types of substances in the past 12 months). Second, to attenuate the sources of bias in labour market participation, we centre on the male population between ages 22 and 50. Limiting the scope to men is relatively common in earlier work (Buchmueller & Zuvekas, 1998; Kandel et al., 1995; Register & Williams, 1992; Ringel et al., 2006; Williams & van Ours, 2020; Zarkin, Mroz et al., 1998;). Below our lower age limit, it is possible that individuals not working are still in education, which is difficult to interpret as a negative economic outcome. Above 50 years old, when, anyway, drug use is quite limited, we run the risk of finding people in early retirement, which is, again, not necessarily a negative situation. Third, to rely on a sufficiently large sample of drug users, we look at drug use in the last 12 months (as opposed to focusing on the last month or any consumption in the lifetime). Fourth, we consider four different labour market outcomes: employment, unemployment, occupational level (holding a white-collar job) and formality. The last variable is proxied by access to health insurance, which is a necessary but not sufficient condition for belonging to the formal labour market. This variable is not available in the second wave considered here. Last, as the number of addictive substances included in the questions is very high, in our heterogeneity analyses, we distinguish between soft and hard drugs, a classification very common in the relevant specialised literature. The former category comprises tranquillisers, cannabis and inhalants, whereas hard addictive substances include opiates, sedatives, stimulants, cocaine, crack, hallucinogens, heroin and methamphetamines. The resulting sample consists of 25,153 observations.

---

[6] The name of the survey in the last wave was changed to the National Survey of Drugs, Alcohol and Tobacco. For simplicity, we use the denomination National Survey of Addictions hereafter.

Table 1 presents the descriptive statistics (means and standard deviations) of the outcome and drug use variables along with those capturing observable characteristics that serve as controls in the econometric analysis (age, ethnicity—whether the interviewed individual speaks an indigenous language—education and area of residence). Note that prime-age Mexican mens' labour market attachment is very high, although less than half work in the formal sector. The share of the male population reporting drug consumption in the last 12 months is 3.2% (2.4 and 1.5% report using soft and hard drugs, respectively).

*Table 1*. Descriptive statistics

|  | Mean | Standard deviation |
|---|---|---|
| Employed (% of working-age population) | 0.969 | 0.172 |
| Unemployed (% of labour force) | 0.028 | 0.165 |
| White-collar worker (% of workers) | 0.344 | 0.475 |
| Formal employment (% of workers) | 0.443 | 0.497 |
| Drug use (last year) | 0.032 | 0.177 |
| Soft drug use (last year) | 0.024 | 0.152 |
| Hard drug use (last year) | 0.015 | 0.123 |
| Age | 34.810 | 8.180 |
| Indigenous | 0.066 | 0.248 |
| Married | 0.719 | 0.449 |
| Less than primary education | 0.096 | 0.295 |
| Primary education | 0.208 | 0.406 |
| Lower secondary education | 0.335 | 0.472 |
| Upper secondary education | 0.224 | 0.417 |
| Higher education | 0.136 | 0.343 |
| Rural area | 0.213 | 0.409 |
| No. of observations | 25,153 | |

*Note*: The descriptive statistics of the all the variables apart from the labour market outcomes refer to adults between 22 and 50 years old who are not in education, disabled or retired. The number of observations for unemployed, white-collar and formal workers is 24,983, 19,168 and 13,787, respectively. Observations are weighted with sampling weights.

*Source:* Authors' analysis from NSA data.

## 2.2. Identification strategy

Establishing a causal relationship between drug use and labour market outcomes requires overcoming several serious endogeneity concerns (Kaestner, 1998; van Ours & Williams, 2015). First, unobserved factors may affect both drug consumption and economic success. In addition, the direction of the bias in ordinary least squares (OLS) estimation due to this problem is not a priori clear. The use of addictive substances might be negatively correlated with parental investment in early childhood or self-control and positively correlated with factors detrimental to labour market prospects such as strong preferences for leisure, a high rate of time preference or consumption of alcohol or tobacco (potentially complementary goods to illicit drugs). If this is the case, OLS would overestimate the negative impact of drug consumption. Nevertheless, the use of illicit addictive substances can correlate with personality traits such as extroversion, sensitivity to peer pressure and risk-taking, factors that contribute to labour market success. Selection on health status (whether those individuals in better or worse health are more likely to consume) may also be an issue.

Second, if drugs are normal goods, the consumption of illicit addictive substances responds positively to income (and thus better labour market performance may lead to higher consumption). The existence of two-way causality in this setting would lead OLS to underestimate the negative impact of drug use.

The last possible source of endogeneity is measurement error. The information on drug use in our study comes from surveys, which are subject to severe underreporting of consumption (Pudney, 2007). When the treatment variable is binary, as in our research design, measurement error leads to attenuation bias (Aigner, 1973); i.e., in such cases, OLS underestimates the negative effect of using illicit addictive substances.

The endogeneity problems outlined above have led researchers to search for appropriate identification strategies to determine the causal effect of drug consumption. Until recently, most of the literature has relied on IV techniques, multiple-equation models with exclusion restrictions or the exploitation of timing of events. In many cases, these papers use instruments that are not very convincing (e.g., religiosity or family background variables, which might exert a direct effect on labour market success), with the few exceptions instead using local substance prices or spatial variation in decriminalisation policies. More recent studies focused on opioids benefit from quasi-experimental research designs that leverage spatial variation in public policies.

Given the lack of a quasi-experimental design, the absence of differences in drug policies across Mexico and the unavailability of local drug prices, we adopt the IV approach proposed by Lewbel (2012), which is also suitable for binary dependent variable models such as ours (Lewbel, 2018). This strategy is appropriate when no convincing external instrument is available, as in our exploration. To the best of our knowledge, we are the first making use of this IV to analyse the impact of drug use.[7] Identification relies on the regressors' being uncorrelated with the product of heteroskedastic errors, a feature by definition present in linear probability models. In the multivariate case, this approach involves the use of several instruments, which allows us to test for overidentifying restrictions using a Sargan–Hansen-type test.

Formally, we aim to estimate the following equation:

$$Y = X'\alpha + D\beta + \varepsilon, \qquad [1]$$

where $Y$ represent a labour market outcome, $D$ denotes drug use, $X$ is a vector that includes an intercept and control variables (age, ethnicity, marital status, education, area of residence and state and year fixed effects) and $\varepsilon$ is an error term. In turn, we assume that drug consumption is a function of the same set of control covariates, i.e.,

$$D = X'\gamma + \upsilon, \qquad [2]$$

where the error term $\upsilon$ may correlate with $\varepsilon$.

Lewbel's approach proceeds in two steps. First, we estimate $\hat{\gamma}$ by OLS, regressing $D$ on $X$, and compute the residuals $\hat{v} = D - X'\hat{\gamma}$. Second, we estimate equation [1] by two-stage least squares (2SLS) using $(Z - \bar{Z}\hat{\gamma})\hat{v}$ as instruments, where $Z$ is a subset of $X$ (or the whole vector $X$, as in our case) and $\bar{Z}$ is the sample mean of $Z$. In this application, $Z$ includes all control covariates. $\hat{\beta}$ estimates the average treatment effect of drug use.

Lewbel's instrument can deal with two sources of bias—unobserved heterogeneity and measurement error—but cannot address eventual reverse causality. Therefore, the IV estimates in our application may underestimate the harm caused by drug consumption, which means that our results represent a lower-bound estimate of the potential negative impact of using illicit addictive substances.

In the robustness checks section, we present additional evidence employing an additional instrument (religiosity), which could solve the problem of two-way causality but cannot adequately address the simultaneity problem and, as we argue there, can at best provide an upper-bound estimate of the impact of drug use. One should be aware the results based on this alternative IV are not fully comparable with those from our main analysis using Lewbel's instruments because both the treatment and the instrument are binary and 2SLS estimates the local average treatment effect, i.e., the effect for those individuals who use drugs because they are not religious, which may represent a relatively modest demographic segment.

We perform all the calculations using the software Stata 18.

---

[7] Applications of this approach in other settings are numerous. See, among many others, Courtemanche et al. (2021) and Islam and Smyth (2015) in health economics and Mallick (2012) and Caliendo et al. (2017) in labour economics.

# 3. Results

## 3.1. Main results

We show the main results of our analysis in Table 2. They include the OLS and IV estimates using the aforementioned Lewbel instruments (which consist of an interaction between the centred regressors and the heteroskedastic residuals of the regression of drug use on the control variables). The instruments are relevant in all cases (the *F*-statistic is well above 10) and Hansen's tests for overidentification do not allow us to reject the null hypotheses in any case.[8]

For all four labour market outcomes, we can reject the null hypothesis that both the OLS and 2SLS coefficients of interest are statistically different from zero. Both sets of estimates have the same sign in all cases, negative in the case of employment, occupational attainment and formality and positive in the case of unemployment. The IV estimates exceed the OLS ones in absolute value, except the IV estimate for formal employment, which is essentially identical to that from OLS. According to the OLS estimates, drug use reduces the employment of Mexican men by 2 percentage points, whereas the IV estimates suggest a larger effect (3.8 points). In the case of unemployment, the OLS regressions imply that drug users are 2.4 percentage points more likely to be unemployed than nonusers, whereas the IV estimates suggest a larger impact (approximately 4 percentage points). Regarding the likelihood of holding a white-collar job (conditional on employment), the OLS and IV estimates yield a negative effect of 4.4 and 6.3 points, respectively. Finally, with respect to formal employment, the negative impact of using illegal drugs amounts to 6.1 percentage points in both cases. These findings are consistent with the priors stated above, particularly the positive selection into drug consumption and the existence of measurement error in the right-hand-side variable. These results also dialogue well with the findings of DeSimone (2002), which, based on external (conventional) credible instruments (state level of depenalisation and local prices of the substance), suggest that OLS underestimates the harm caused by marijuana use. Interestingly, these results suggest worse consequences of drug use for labour market success than those reported for high-income countries. In the rest of this section (examining the heterogeneity and robustness of our results), we centre on and show only the IV estimates of the effect.

---

[8] In contrast to the estimation using external (conventional) instruments, the details of the first stage here are not intrinsically interesting beyond the tests for joint significance (*F*-statistic) and overidentifying restrictions (Hansen's *J* test).

*Table 2*. Effects of drug use on labour market outcomes

| | Employment | | Unemployment | | White-collar occupation | | Formal employment | |
|---|---|---|---|---|---|---|---|---|
| | OLS | IV | OLS | IV | OLS | IV | OLS | IV |
| | (I) | (II) | (III) | (VI) | (V) | (VI) | (VII) | (VIII) |
| Drug use | −0.020** | −0.038*** | 0.024*** | 0.039*** | −0.044*** | −0.063** | −0.061** | −0.061** |
| | (0.008) | (0.014) | (0.008) | (0.014) | (0.016) | (0.025) | (0.024) | (0.026) |
| Adjusted R$^2$ | 0.015 | | 0.014 | | 0.305 | | 0.183 | |
| No. of observations | 25,153 | 25,153 | 24,983 | 24,983 | 19,168 | 19,168 | 13,787 | 13,787 |
| Mean of dependent variable | 0.967 | 0.967 | 0.028 | 0.028 | 0.317 | 0,317 | 0.408 | 0.408 |
| Mean of independent variable | 0.033 | 0.033 | 0.034 | 0.034 | 0.032 | 0.032 | 0.028 | 0.028 |
| First-stage *F*-statistic | | 40.546 | | 39.933 | | 29.395 | | 365.411 |
| Hansen *J* test (*p*-value) | | 0.396 | | 0.385 | | 0.719 | | 0.473 |
| Control variables | ✓ | ✓ | ✓ | ✓ | ✓ | ✓ | ✓ | ✓ |
| State fixed effects | ✓ | ✓ | ✓ | ✓ | ✓ | ✓ | ✓ | ✓ |
| Year fixed effects | ✓ | ✓ | ✓ | ✓ | ✓ | ✓ | ✓ | ✓ |

*Notes*: *** significant at 1% level; ** significant at 5% level; * significant at 10% level. The table shows the 2SLS results obtained with Lewbel's instruments. The control variables are age, age squared, ethnicity, marital status, education and area of residence. Heteroskedasticity-robust standard errors are in parentheses.

Source: Authors' analysis.

## 3.2. Heterogeneity analysis

We examine the heterogeneity in our results across different demographic groups (Tables 3–5). One should be cautious in interpreting these results because of the reduction in sample size and hence statistical power when we consider the demographic subsamples. First, regarding age (Table 3), the impact of drug use seems to be larger on young Mexicans than on those aged more than 35 years in terms of employment and unemployment. The opposite is the case for the likelihood of being a white-collar worker. The estimated coefficients for several subsamples are not statistically different from zero. Moreover, we cannot reject the null hypothesis in the overidentification test for the occupational attainment of the eldest group, so this result should be interpreted with particular care.

Second, with respect to educational level (Table 4), the estimated coefficients for low-educated individuals are statistically different from zero in all cases. This is not the case for Mexicans with a high level of schooling. Nevertheless, we cannot rule out that the difference between the two groups is zero in any of the dimensions analysed.

Finally, we test whether the impact of hard drugs is greater than that of the soft substances (Table 5). We find that the size of the coefficients is always larger in the former group of drugs, which is reassuring, even though we cannot reject that their effect is statistically different from that of the soft ones.

Overall, our heterogeneity analysis suggests that the impact seems to be more relevant for young and poorly educated men and in the case of hard drugs. These results align well with those in the previous literature on drug use.

*Table 3*. Effects of drug use on labour market outcomes by age group (IV estimates)

|  | Employment | | Unemployment | | White-collar occupation | | Formal employment | |
|---|---|---|---|---|---|---|---|---|
|  | 22–35 years old | 36–50 years old | 22–35 years old | 36–50 years old | 22–35 years old | 36–50 years old | 22–35 years old | 36–50 years old |
|  | (I) | (II) | (III) | (VI) | (V) | (VI) | (VII) | (VIII) |
| Drug use | −0.054*** | 0.015 | 0.048** | −0.005 | −0.011 | −0.095** | −0.083** | −0.015 |
|  | (0.019) | (0.013) | (0.019) | (0.013) | (0.033) | (0.025) | (0.032) | (0.041) |
| Difference | −0.069*** | | 0.054*** | | 0.084** | | −0.068 | |
|  | (0.023) | | (0.023) | | (0.041) | | (0.052) | |
| No. of observations | 12,418 | 12,735 | 12,321 | 12,662 | 9,439 | 9,729 | 6,973 | 6,814 |
| Mean of dependent variable | 0.962 | 0.973 | 0.032 | 0.025 | 0.313 | 0.320 | 0.408 | 0.409 |
| Mean of independent variable | 0.043 | 0.024 | 0.043 | 0.025 | 0.040 | 0.024 | 0.034 | 0.021 |
| First-stage *F*-statistic | 22.563 | 264.091 | 21.916 | 260.136 | 18.378 | 133.779 | 953.310 | 1,810.296 |
| Hansen *J* test (*p*-value) | 0.495 | 0.187 | 0.528 | 0.186 | 0.871 | 0.009 | 0.112 | 0.480 |
| Control variables | ✓ | ✓ | ✓ | ✓ | ✓ | ✓ | ✓ | ✓ |
| State fixed effects | ✓ | ✓ | ✓ | ✓ | ✓ | ✓ | ✓ | ✓ |
| Year fixed effects | ✓ | ✓ | ✓ | ✓ | ✓ | ✓ | ✓ | ✓ |

*Notes*: *** significant at 1% level; ** significant at 5% level; * significant at 10% level. The table shows the 2SLS results obtained with Lewbel's instruments. The control variables are age, age squared, ethnicity, marital status, education and area of residence. Heteroskedasticity-robust standard errors are in parentheses.

Source: Authors' analysis.

*Table 4.* Effects of drug use on labour market outcomes by educational attainment (IV estimates)

| | Employment | | Unemployment | | White-collar occupation | | Formal employment | |
|---|---|---|---|---|---|---|---|---|
| | Low education | High education | Low education | High education | Low education | High education | Low education | High education |
| | (I) | (II) | (III) | (VI) | (V) | (VI) | (VII) | (VIII) |
| Drug use | −0.037 | −0.029 | 0.039** | 0.027 | −0.055** | −0.026 | −0.058** | −0.067 |
| | (0.016) | (0.019) | (0.016) | (0.018) | (0.028) | (0.043) | (0.029) | (0.055) |
| Difference | −0.007 | | 0.012 | | −0.029 | | 0.009 | |
| | (0.025) | | (0.025) | | (0.051) | | (0.062) | |
| No. of observations | 16,974 | 8,179 | 16,852 | 8,131 | 13,053 | 6,115 | 9,501 | 4,286 |
| Mean of dependent variable | 0.965 | 0.972 | 0.030 | 0.025 | 0.173 | 0.624 | 0.325 | 0.592 |
| Mean of independent variable | 0.025 | 0.015 | 0.025 | 0.015 | 0.024 | 0.014 | 0.020 | 0.013 |
| First-stage $F$-statistic | 47.972 | 28.291 | 47.349 | 28.078 | 32.051 | 27.504 | 1,344.067 | 939.008 |
| Hansen $J$ test ($p$-value) | 0.445 | 0.594 | 0.486 | 0.582 | 0.733 | 0.159 | 0.540 | 0.907 |
| Control variables | ✓ | ✓ | ✓ | ✓ | ✓ | ✓ | ✓ | ✓ |
| State fixed effects | ✓ | ✓ | ✓ | ✓ | ✓ | ✓ | ✓ | ✓ |
| Year fixed effects | ✓ | ✓ | ✓ | ✓ | ✓ | ✓ | ✓ | ✓ |

*Notes*: *** significant at 1% level; ** significant at 5% level; * significant at 10% level. The table shows the 2SLS results obtained with Lewbel's instruments. The control variables are age, age squared, ethnicity, marital status, education and area of residence. Heteroskedasticity-robust standard errors are in parentheses.

Source: Authors' analysis.

*Table 5.* Effects of drug use on labour market outcomes by type of drug (IV estimates)

| | Employment | | Unemployment | | White-collar occupation | | Formal employment | |
|---|---|---|---|---|---|---|---|---|
| | Soft drug | Hard drug | Soft drug | Hard drug | Soft drug | Hard drug | Soft drug | Hard drug |
| | (I) | (II) | (III) | (VI) | (V) | (VI) | (VII) | (VIII) |
| Drug use | −0.025* | −0.052** | 0.030** | 0.044** | −0.037 | −0.069** | −0.039 | −0.049 |
| | (0.014) | (0.021) | (0.014) | (0.020) | (0.024) | (0.033) | (0.029) | (0.036) |
| Difference | 0.027 | | −0.013 | | 0.032 | | 0.010 | |
| | (0.022) | | (0.022) | | (0.035) | | (0.037) | |
| No. of observations | 25,143 | 25,149 | 24,973 | 24,979 | 19,159 | 19,165 | 13,783 | 13,783 |
| Mean of dependent variable | 0.967 | 0.967 | 0.028 | 0.028 | 0.317 | 0.317 | 0.408 | 0.408 |
| Mean of independent variable | 0.025 | 0.015 | 0.025 | 0.015 | 0.024 | 0.014 | 0.020 | 0.013 |
| First-stage $F$-statistic | 47.972 | 28.291 | 47.349 | 28.078 | 32.051 | 27.504 | 1,344.067 | 939.008 |
| Hansen $J$ test ($p$-value) | 0.445 | 0.594 | 0.486 | 0.582 | 0.733 | 0.159 | 0.540 | 0.907 |
| Control variables | ✓ | ✓ | ✓ | ✓ | ✓ | ✓ | ✓ | ✓ |
| State fixed effects | ✓ | ✓ | ✓ | ✓ | ✓ | ✓ | ✓ | ✓ |
| Year fixed effects | ✓ | ✓ | ✓ | ✓ | ✓ | ✓ | ✓ | ✓ |

*Notes*: *** significant at 1% level; ** significant at 5% level; * significant at 10% level. The table shows the 2SLS results obtained with Lewbel's instruments. The control variables are age, age squared, ethnicity, marital status, education and area of residence. Heteroskedasticity-robust standard errors are in parentheses.

Source: Authors' analysis.

## 3.3. Robustness checks

The final part of this section discusses the results of several robustness checks aimed at assessing the stability of the findings of our main analysis. First, we look at the outcomes when we use an external instrument, namely, whether the person is religious (which in Mexico is essentially synonymous with being a Catholic Christian). It is well established that devout people tend to make lower use of addictive substances (Koenig et al., 2023). Earlier literature has widely made use of different variations of the religiosity IV (see, e.g., Kaestner [1991], MacDonald and Pudney [2000a, 2001], Register and Williams [1992] and Roebuck et al. [2004]). In principle, this instrument could address the possible reverse causality between drug use and labour market outcomes. Nevertheless, it is not exempt from problems: being religious might have a direct effect on labour market outcomes through different channels such as social capital (Iyer, 2016; Maselko et al., 2011). For instance, pious men could have a better support network than others. Given that most Mexicans are Catholic and the survey does not allow us to distinguish the degree of religiosity, it is reasonable to interpret this variable as nonextremist religiosity, which leads us to expect a positive effect of the variable on economic outcomes.[10] If our expectation is correct, the IV estimates will overestimate the impact of drugs on the labour market. Therefore, we consider the results based on this IV upper-bound estimates of the impact of drug use. Table A1 shows the results of this analysis, which indicate that the instrument is not weak and its estimated coefficient in the first stage has the expected (negative) sign and which, importantly, are coherent with our interpretation above. Overall, the estimated effect is also negative and larger (but not statistically different from zero in the case of occupational attainment) and much less precise (with a larger standard error) than that obtained with our preferred set of instruments (the Lewbel IVs).

Second, we repeat our main exploration excluding marital status from the regressors (Table A2), which can raise additional endogeneity concerns in the case of drug use (e.g., drug users could have a lower chance of finding a partner or a greater likelihood of splitting up or having a lower educational attainment). These estimates are not statistically different from those reported in Section 3.1, suggesting that the main harm caused by drugs is direct (rather than mediated by education or marital status).[11]

Our final robustness exercise (Table A3) tests an alternative taxonomy of drugs (recreational vs. dependency drugs, with the latter being potentially more harmful than the former) used by MacDonald and Pudney (2000a).[12] The results are quite similar to those comparing soft and hard drug users (dependency substances appear to exert a larger negative effect than recreational ones, but the point estimates are not statistically different), which is reassuring.

---

[10] For example, the evidence suggests the existence of a Catholic wage premium for men in countries where this religion is the majority one, such as the United States (Mohanty, 2023), Canada (1985), Germany (Sinnewe et al., 2016), Australia (Kort & Dollery, 2012) and Brazil (Bernardelli et al., 2020).
[11] Note that we reject the null hypothesis that all the overidentifying restrictions are valid at the 10% significance level for occupational attainment, so this result should be interpreted with some additional caution.

## 4. Conclusion

Over the past two decades, Mexico has, regrettably, become notorious for the public safety problems associated with cartel activity, but evidence on the consequences of becoming a drug supplier and hub for narcotrafficking for other dimensions of citizens' lives is limited. This paper has explored the impact of drug consumption on the labour market outcomes of prime-age Mexican men and represents the first such empirical study for a middle-income economy. We have also illustrated the use of Lewbel's IV approach in the context of drug use, where the absence of credible instruments is the norm.

Our results suggest that the use of addictive substances severely harms men's labour market prospects, not only by reducing their participation but also by affecting their occupational attainment and formality. These effects appear to be more relevant for hard drugs, although these results are subject to limitations because of small subsample sizes. Interestingly, even though the comparability across different studies is limited, overall, these results are somewhat more negative than those observed for highly developed countries. Among the causes that could explain this result, we speculate that it could be attributable to budgetary constraints that affect antidrug campaigns and addiction treatment and weaker social safety nets (García, 2024; Rafful, 2020). Further research could attempt to disentangle the role of these factors.

---

[12] Recreational drugs include stimulants, cannabis, hallucinogens, inhalants and methamphetamines, whereas dependency addictive substances comprise opiates, cocaine, crack, heroin, tranquilizers and sedatives.

# Annex

*Table A1.* Effects of drug use on labour market outcomes using religiosity as instrument

|  | Employment | Unemployment | White-collar occupation | Formal employment |
|---|---|---|---|---|
|  | (I) | (II) | (III) | (VI) |
| Drug use | −0.279** | 0.280** | −0.141 | −0.807** |
|  | (0.125) | (0.128) | (0.323) | (0.406) |
| No. of observations | 25,153 | 24,983 | 19,168 | 13,787 |
| Mean of dependent variable | 0.967 | 0.028 | 0.317 | 0.408 |
| Mean of independent variable | 0.033 | 0.034 | 0.032 | 0.028 |
| Mean of the instrument | 0.907 | 0.907 | 0.911 | 0.924 |
| First-stage estimate of religiosity | −0.030*** | −0.035*** | −0.035*** | −0.039*** |
|  | (0.005) | (0.005) | (0.006) | (0.008) |
| First-stage *F*-statistic | 41.071 | 41.746 | 24.332 | 24.434 |
| Control variables | ✓ | ✓ | ✓ | ✓ |
| State fixed effects | ✓ | ✓ | ✓ | ✓ |
| Year fixed effects | ✓ | ✓ | ✓ | ✓ |

*Notes*: *** significant at 1% level; ** significant at 5% level; * significant at 10% level. The table shows the 2SLS results obtained with religiosity as instrument. The control variables are age, age squared, ethnicity, marital status, education and area of residence. Heteroskedasticity-robust standard errors are in parentheses. The first-stage estimate of religiosity is the estimated coefficient of this variable in the first-stage regression, which includes the same set of controls and fixed effects as the equation estimated in the table.

Source: Authors' analysis.

*Table A2.* Effects of drug use on labour market outcomes (IV results excluding marital status and education)

|  | Employment | Unemployment | White-collar occupation | Formal employment |
|---|---|---|---|---|
|  | (I) | (II) | (III) | (VI) |
| Drug use | −0.041*** | 0.041*** | −0.122*** | −0.127*** |
|  | (0.014) | (0.014) | (0.013) | (0.025) |
| No. of observations | 25,153 | 24,983 | 19,168 | 13,787 |
| Mean of dependent variable | 0.967 | 0.028 | 0.317 | 0.408 |
| Mean of independent variable | 0.033 | 0.034 | 0.032 | 0.028 |
| First-stage $F$-statistic | 359.321 | 353.859 | 244.119 | 413.129 |
| Hansen $J$ test ($p$-value) | 0.253 | 0.168 | 0.066 | 0.174 |
| Control variables | ✓ | ✓ | ✓ | ✓ |
| State fixed effects | ✓ | ✓ | ✓ | ✓ |
| Year fixed effects | ✓ | ✓ | ✓ | ✓ |

*Notes*: *** significant at 1% level; ** significant at 5% level; * significant at 10% level. The table shows the 2SLS results obtained with Lewbel's instruments. The control variables are age, age squared, ethnicity, education and area of residence. Heteroskedasticity-robust standard errors are in parentheses.

Source: Authors' analysis.

*Table A3.* Effects of drug use on labour market outcomes by type of drug (recreational vs. dependency drugs, IV estimates)

| | Employment | | Unemployment | | White-collar occupation | | Formal employment | |
|---|---|---|---|---|---|---|---|---|
| | Recreational drug | Dependency drug | Recreational drug | Dependency drug | Recreational drug | Dependency drug | Recreational drug | Dependency drug |
| | (I) | (II) | (III) | (VI) | (V) | (VI) | (VII) | (VIII) |
| Drug use | −0.034** | −0.043** | 0.036** | 0.044** | −0.043** | −0.067* | −0.031 | −0.043 |
| | (0.015) | (0.021) | (0.015) | (0.021) | (0.025) | (0.037) | (0.030) | (0.036) |
| Difference | 0.009 | | −0.008 | | 0.013 | | 0.012 | |
| | (0.022) | | (0.022) | | (0.039) | | (0.036) | |
| No. of observations | 25,144 | 25,148 | 24,974 | 24,978 | 19,160 | 19,164 | 13,779 | 13,785 |
| Mean of dependent variable | 0.967 | 0.967 | 0.028 | 0.028 | 0.317 | 0.317 | 0.408 | 0.408 |
| Mean of independent variable | 0.026 | 0.015 | 0.026 | 0.015 | 0.025 | 0.013 | 0.020 | 0.013 |
| First-stage $F$-statistic | 53.179 | 19.146 | 52.386 | 18.922 | 36.856 | 16.282 | 993.073 | 878.055 |
| Hansen $J$ test ($p$-value) | 0.565 | 0.684 | 0.567 | 0.772 | 0.636 | 0.129 | 0.482 | 0.847 |
| Control variables | ✓ | ✓ | ✓ | ✓ | ✓ | ✓ | ✓ | ✓ |
| State fixed effects | ✓ | ✓ | ✓ | ✓ | ✓ | ✓ | ✓ | ✓ |
| Year fixed effects | ✓ | ✓ | ✓ | ✓ | ✓ | ✓ | ✓ | ✓ |

*Notes*: *** significant at 1% level; ** significant at 5% level; * significant at 10% level. The table shows the 2SLS results obtained with Lewbel's instruments. The control variables are age, age squared, ethnicity, marital status, education and area of residence. Heteroskedasticity-robust standard errors are in parentheses.

Source: Authors' analysis.